
\input harvmac

\overfullrule=0pt



\def\bs{\bigskip}
\def\no{\noindent}
\def\hb{\hfill\break}
\def\qq{\qquad}
\def\bl{\bigl}
\def\br{\bigr}

\def\IR{\relax{\rm I\kern-.18em R}}

\def\slsu{$SL(2,\IR)_{-k'}\otimes SU(2)_k/(\IR \otimes \tilde \IR)$}

\def\np {  Nucl. Phys. }


\def\k{\kappa}
\def\r{\rho}
\def\a{\alpha}

\def\b{\beta}

\def\d{\delta}

\def\th{\theta}

\def\m{\mu}
\def\n{\nu}

\def\l{\lambda}

\def\IR{\relax{\rm I\kern-.18em R}}
\def\gh{$SL(2,\IR)\otimes SO(1,1)^{d-2}/SO(1,1)$}
\def\pap{\partial_+}
\def\pam{\partial_-}
\def\papm{\partial_{\pm}}
\def\pat{\partial_{\tau}}



hep-th/9301047 \hfill {USC-93/HEP-B1}
\rightline{January 1993}
\bs\bs

\centerline  {\bf  EXACT EFFECTIVE ACTION AND SPACETIME GEOMETRY}
\centerline   {   {\bf IN GAUGED WZW MODELS }
                  {  \footnote{$^\dagger$}
 {Research supported in part by DOE, under Grant No. DE-FG03-84ER-40168. } }
               }

\vskip 1.00 true cm

\centerline {I. BARS and K. SFETSOS}

\bigskip

\centerline {Physics Department}
\centerline {University of Southern California}
\centerline {Los Angeles, CA 90089-0484, USA}


\vskip 1.50 true cm

\centerline{ABSTRACT}

\bs

We present an effective quantum action for the gauged WZW model
$G_{-k}/H_{-k}$. It is conjectured that it is valid to all orders of the
central extension $(-k)$ on the basis that it reproduces the exact spacetime
geometry of the zero modes that was previously derived in the algebraic
Hamiltonian formalism. Besides the metric and dilaton, the new results that
follow from this approach include the exact axion field and the solution of
the geodesics in the exact geometry. It is found that the axion field is
generally non-zero at higher orders of $1/k$ even if it vanishes at large $k$.
We work out the details in two specific coset models, one non-abelian, i.e.
$SO(2,2)/SO(2,1)$ and one abelian, i.e $SL(2,\IR)\otimes SO(1,1)^{d-
2}/SO(1,1)$. The simplest case $SL(2,\IR)/\IR$ corresponds to a limit.


 \lref\HOHOS{ J. H. Horne, G. T. Horowitz and A. R. Steif, Phys. Rev. Lett.
{\bf 68} (1991) 568.}
 \lref\IBhet{ I. Bars, Nucl. Phys. {\bf B334} (1990) 125. }
 \lref\BN{ I. Bars and D. Nemeschansky, Nucl. Phys. {\bf B348} (1991) 89.}
 \lref\IBCS{ I. Bars,``Curved Spacetime Strings and Black Holes", in Proc.
 {\it XX$^{th}$ Int. Conf. on Diff. Geometrical Methods in Physics}, eds. S.
Catto and A. Rocha, Vol. 2, p. 695, (World Scientific, 1992).}
 \lref\WIT{E. Witten, Phys. Rev. {\bf D44} (1991) 314.}.
 \lref\CRE{M. Crescimanno, Mod. Phys. Lett. {\bf A7} (1992) 489.}
 \lref\HOHO{J. B. Horne and G. T. Horowitz, Nucl. Phys. {\bf B368} (1992) 444.}
 \lref\BSthree{I. Bars and K. Sfetsos, Mod. Phys. Lett. {\bf A7} (1992) 1091.}
 \lref\BShet{I. Bars and K. Sfetsos, Phys. Lett. {\bf 277B} (1992) 269.}
 \lref\FRA{E. S. Fradkin and V. Ya. Linetsky, Phys. Lett. {\bf 277B}
          (1992) 73.}
 \lref\ISH{N. Ishibashi, M. Li, and A. R. Steif,
         Phys. Rev. Lett. {\bf 67} (1991) 3336.}
 \lref\HOR{P. Horava, Phys. Lett. {\bf 278B} (1992) 101.}
 \lref\RAI{E. Raiten, ``Perturbations of a Stringy Black Hole'',
         Fermilab-Pub 91-338-T.}
 \lref\GER{D. Gershon, ``Exact Solutions of Four-Dimensional Black Holes in
         String Theory'', TAUP-1937-91.}
 \lref \GIN {P. Ginsparg and F. Quevedo,  Nucl. Phys. {\bf B385} (1992) 527. }
 \lref\BSglo{I. Bars and K. Sfetsos, Phys. Rev. {\bf D46} (1992) 4495.}

 \lref\BSexa{I. Bars and K. Sfetsos, Phys. Rev. {\bf D46} (1992) 4510.}

 \lref\SFET{K. Sfetsos, ``Conformally Exact Results for
$SL(2,\IR)\otimes SO(1,1)^{d-2} /SO(1,1)$ Coset Models'',
USC-92/HEP-S1 (hep-th/9206048), to appear in Nucl. Phys. {\bf B}.}

 \lref\groups{
 M. Crescimanno. Mod. Phys. Lett. {\bf A7} (1992) 489. \hb
 J. B. Horne and G.T. Horowitz, Nucl. Phys. {\bf B368} (1992) 444. \hb
 E. S. Fradkin and V. Ya. Linetsky, Phys. Lett. {\bf 277B} (1992) 73. \hb
 P. Horava, Phys. Lett. {\bf 278B} (1992) 101.\hb
 E. Raiten, ``Perturbations of a Stringy Black Hole'',
         Fermilab-Pub 91-338-T.\hb
 D. Gershon, ``Exact Solutions of Four-Dimensional Black Holes in
         String Theory'', TAUP-1937-91.}

 \lref\KASU{Y. Kazama and H. Suzuki, Nucl. Phys. {\bf B234} (1989) 232 \semi
Phys. Lett. {\bf 216 B} (1989) 112.}

 \lref\NAWIT{C. Nappi and E. Witten, Phys. Lett. {\bf 293B} (1992) 309.}

 \lref\BSslsu{I. Bars and K. Sfetsos
$SL(2,\IR)\otimes SU(2)/\IR^2$ String Model
in Curved Spacetime and Exact Conformal Results", USC-92/HEP-B3 (hep-th/
9208001), to appear in Phys. Lett.{\bf B}.}

 \lref\WITanom{E. Witten, Comm. Math. Phys. {\bf 144} (1992) 189.}

 \lref\IBhetero{I. Bars, Phys. Lett. {\bf 293B} (1992) 315.}
 \lref\IBerice{I. Bars, ``Superstrings on Curved Spacetimes", USC-92/HEP-B5
(hep-th 9210079). }

 \lref\DVV{R. Dijkgraaf, E. Verlinde and H. Verlinde, Nucl. Phys. {\bf B371}
(1992) 269.}

 \lref\TSEY{A.A. Tseytlin, Phys. Lett. {\bf 268B} (1991) 175.
      \semi I. Jack, D. R. T. Jones and J. Panvel, ``Exact Bosonic and
Supersymmetric String Black Hole Solutions", LTH-277.}.

 \lref\BST { I. Bars,  K. Sfetsos and A.A. Tseytlin, unpublished. }
 \lref\TSEYT{ A.A. Tseytlin, ``Effective Action in Gauged WZW Models and Exact
String Solutions", Imperial/TP/92-93/10.}

 \lref\SHIF { M.A. Shifman, Nucl. Phys. {\bf B352} (1991) 87.}
 \lref\SHIFM { H. Leutwyler and M.A. Shifman,
                    Int. J. Mod. Phys. {\bf A7} (1992) 795. }
 \lref\POLWIG { A.M. Polyakov and P.B. Wiegman, Phys. Lett. {\bf 131B} (1984)
                             121.  }
 \lref\BCR{K. Bardakci, M. Crescimanno and E. Rabinovoici,
               Nucl. Phys. {\bf B344} (1990) 344. }
 \lref\GWZW{ E. Witten, \np {\bf B223} (1983)422.
         \semi   K. Bardakci, E. Rabinovici andB. S\"aring,
                        Nucl. Phys. {\bf B299} (1988) 157.
         \semi K. Gawedzki and A. Kupiainen, Phys. Lett. {\bf 215B} (1988) 119;
             Nucl. Phys. {\bf B320} (1989) 625. }
 \lref\SCH{ D. Karabali, Q-Han Park, H.J. Schnitzer and Z.Yang,
                   Phys. Lett. {\bf B216} (1989) 307.
    \semi  D. Karabali and H.J. Schnitzer, Nucl. Phys. {\bf B329} (1990) 649. }

 \lref\KIR{E. Kiritsis, Mod. Phys. Lett. {\bf A6} (1991) 2871. }


\vfill
\eject


\newsec {Introduction}

A gauged WZW model can be rewritten in the form of a non-linear sigma model by
choosing a unitary gauge that eliminates some of the degrees of freedom from
the group element, and then integrating out the non-propagating gauge fields
\BN\WIT . The remaining degrees of freedom are identified with string
coordinates $X^\mu(\tau,\sigma)$.
The resulting action exhibits a gravitational metric $G_{\mu\nu}(X)$ and an
antisymmetric tensor $B_{\mu\nu}(X)$ at the classical level. At the one loop
level there is also a dilaton $\Phi(X)$. These fields govern the spacetime
geometry of the manifold on which the string propagates. Conformal invariance
at one loop level demands that they satisfy coupled Einstein's equations.
Thanks to the exact conformal properties of the gauged WZW model these
equations are automatically satisfied.

For a restricted list of non-compact gauged WZW models there is only one time
coordinate \BN\IBCS\GIN , thus making them suitable for a string theory
interpretation in curved spacetime. The list may be extended to supersymmetric
heterotic models \BShet\IBhetero\IBerice . Then these models can be viewed as
generating automatically a solution of these rather unyielding Einstein
equations. One only needs to do some straightforward algebra based on group
theory to extract the explicit forms of $G_{\mu\nu}, B_{\mu\nu}, \Phi$.
Following the lead of \WIT\ who interpreted the $SL(2,\IR)_{-k}/\IR$ case at
$k=9/4$ \BN\ as a string propagating in the background geometry of a black
hole in two dimensions, several groups have worked out the geometry for all
possible cases up to dimension four \BSthree\groups\BShet\NAWIT . The
resulting new geometries are generally non-isotropic and have singularities
that are more intricate than a black hole, and may have physical
interpretations in the early string universe. The global aspects of these
higher geometries have been understood \BSglo\BSslsu\IBerice . They all have
very interesting duality properties that correspond to interchanges of patches
of the global geometry. This duality may be viewed as inversions in group
space \BSglo\ and are related to asymmetric left-right gauging that involves a
twist on the right relative to the left of the group element \BSthree .

Since these are singular geometries it is clearly desirable to go beyond the
one loop expansion of the effective sigma model and consider the effect of
the exact conformal invariance that underlies the gauged WZW model.
It is the purpose of the present paper to accomplish this by considering the
full quantum effective action. Of course, the full quantum action is of
interest in its own right since the range of its applications goes far beyond
the exact geometry of the model. However, at this point, rather than a
complete derivation we are able to present a conjecture on the form of the
full quantum effective action. We will justify its form by deriving the exact
geometry and comparing to our previous exact results obtained with algebraic
Hamiltonian techniques. Therefore, let us first briefly review the status of
conformally exact results.

In recent papers \BSexa\SFET\BSslsu\ we showed how to improve on the
perturbative Lagrangian results by using algebraic Hamiltonian techniques to
compute globally valid and conformally exact geometrical quantities such as
the metric and dilaton (and, in principle, other fields) in gauged WZW models.
We have applied the method to bosonic, heterotic and type-II supersymmetric 4D
string models that use the non-compact cosets. The main idea is as follows. It
is part of the folklore of string theory that $L_0+\bar L_0$ is the Laplacian,
and that when applied to the tachyon $T$ it takes the form

\eqn\laplacian{ (L_0+\bar L_0)T={-1\over e^\Phi\sqrt{-
G}}\partial_\mu(e^\Phi\sqrt{-G}G^{\mu\nu}\partial_\nu T) \ .}
This equation follows from the general form of the low energy effective action
of string theory which concentrates on the low lying spectrum. Eq.\laplacian\
was used in \DVV\ where the $SL(2,\IR)_{-k}/\IR$ geometry to all orders in
$1/k$ was ``conjectured" to arise from it. Indeed this simplest case has been
checked to work up to four loops for the bosonic string and up to five loops
for the type-II superstring \TSEY . In \BSexa\ we developed the general
methods to use \laplacian\ to extract the {\it global} and conformally {\it
exact} geometry for all $G/H$ models, including the heterotic superstring
case. This was based on the following proof of \laplacian\ which was implicit
but was not stated explicitly in \BSexa : Evidently, $L_0+\bar L_0$, as
constructed from currents in a $G/H$ theory, is exact to all orders in $1/k$.
The tachyon is annihilated by all $n\ge 1$ currents $J_n^G$, so that only the
zero mode currents $J_0^G$ are relevant, as they appear in $L_0+\bar L_0$. We
further made the reasonable assumption that the tachyon wavefunction depends
only on the zero modes of the group parameters. Therefore, we only need to
know how to construct the zero mode currents from the zero modes of the group
parameters as differential operators. We have shown in \BSexa\ how to
accomplish this, so that $L_0+\bar L_0$ becomes a second order differential
operator. Then, after using the crucial observation that the tachyon $T$ is
constructed from certain gauge invariant combinations of group parameters, and
then applying the chain rule as described in \BSexa , $L_0+\bar L_0 $ indeed
takes the general form of the Laplacian in \laplacian\ with a non-trivial
dilaton and metric. Since this Laplacian is exact to all orders in $1/k$ the
resulting metric and dilaton must be identified with the exact ones to all
orders in $1/k$.

The Hamiltonian approach has effectively concentrated on the zero modes.
Therefore, in comparing the old results to the new exact quantum action of
the present paper, we must take care that the exact geometry is in agreement
first and foremost for the zero modes. We shall see that the geometry  for the
higher modes may be non-local on the world sheet. Our approach here will apply
to abelian cosets such as $SL(2,\IR)/\IR$, \gh\ or \slsu\ as well as
non-abelian
ones such as $SO(2,2)/SO(2,1)\sim SL(2,\IR)\otimes SL(2,\IR)/SL(2,\IR )$ or
$SO(3,2)/SO(3,1)$. For related results for the abelian coset $SL(2,\IR)/\IR$
see also a paper by A.A. Tseytlin \TSEYT\ with whom the present investigations
were initiated \BST .


\newsec {The effective quantum action}

The effective quantum action for any field theory is derived by introducing
sources and then applying a Legendre transform
 \ref\ITZ{see e.g. C. Itzykson and J. Zuber, {\it Quantum Field Theory},
McGraw Hill (1980). }.
 The effective action, which is then used as a classical field theory,
incorporates all the higher loop effects. Based on a perturbative analysis in
\SHIF\SHIFM\ it has been argued \TSEYT\ that for the
ungauged WZW model $G_{-k}$ this procedure gives

\eqn\ewzw{ \eqalign {
  & S^{eff}_{WZW} = (-k+g)I_0(g), \cr
  & I_0(g) ={1\over 8\pi}\int_M Tr(\pap g^{-1}\pam g)+{1\over 24\pi}
  \int_B Tr(g^{-1}dg)^3    \ ,     }}
Therefore the full quantum effective action differs from the classical one
only by the overall renormalization that replaces $(-k)$ by $(-k+g)$, where
$g$ is the Coxeter number for the group $G$, not to be confused with the group
element $g(\sigma^+,\sigma^-)$ (we have also assumed a conformally critical
theory with the Virasoro central charge at $c=26$ that fixes the value of
$k$). Instead of relying on the perturbative approach in \SHIF\SHIFM\TSEYT\
we can justify the result \ewzw\ by the following argument on the geometry:
Before the quantum effects are taken into account the classical sigma model
geometry of the WZW model is given by the group manifold metric and the
antisymmetric tensor (the axion), both {\it multiplied by $(-k)$}. To derive
the exact geometry by the algebraic Hamiltonian approach one must use the
quantum exact stress tensor to construct $L_0+\bar L_0$ as described in
section-1.  The conformally exact quantum stress tensor follows from the
classical one by a well known renormalization that replaces $(-k)$ by $(-
k+g)$. It follows from this that the exact geometry in the Hamiltonian
approach is the same as the classical geometry except for the aforementioned
renormalization. To agree with this quantum result the exact effective action
must be the same as the classical one except for the proportionality constant
$(-k+g)$ as given in \ewzw . Furthermore, $g(\sigma^+,\sigma^-)$ is now
treated as a classical field.

We now extend these arguments to the gauged WZW model (GWZW) for $G_{-k}/H_{-
k}$ which is defined by the classical action  \GWZW \SCH

\eqn\agwzw{ \eqalign {
     & S_{GWZW}=-k\ I_0(g)-k\ I_1(g,A_+,A_-)\ , \cr
     & I_1(g,A_+,A_-)={1\over 4\pi}\int_M Tr(A_- \pap g g^{-1}
-A_+ g^{-1}\pam g +A_-gA_+g^{-1}-A_+A_-)\ .\cr} }
Here $g$ is a group element in $G$ and $A_{\pm}$ is valued in the Lie algebra
for the subgroup $H$. This action is invariant under the local gauge
transformations that belong to the subgroup H
 \foot {The more general left-right asymmetric gauging of \BSthree\ may also
be discussed in a straightforward fashion (for an application see \BSslsu\ ).}

\eqn\gaugetr{ g\to \Lambda^{-1}g\Lambda ,\qquad A_{\pm}\to \Lambda^{-
1}(A_{\pm}-\partial_{\pm})\Lambda \ .}
It is useful to make a change of variables to group elements $h_{\pm}\in H$,
$A_+=\partial_+h_+h_+^{-1},\ A_-=\partial_-h_-h_-^{-1}$. After picking up a
determinant and an anomaly from the measure, the path integral
is rewritten with a new form for the action \POLWIG\SCH

\eqn\agroup{ S_{GWZW}=-kI_0(h_-^{-1}gh_+)+(k-2h)I_0(h_-^{-1}h_+) \ ,}
which is manifestly gauge invariant under $h_\pm\to \Lambda^{-1}h_\pm$.
The new path integral measure is the
Haar group measure  $Dg\ Dh_+\ Dh_-$. We want to take advantage of the
similarity of this action to the classical WZW action: the first term is
appropriate for $G$ with central extension $(-k)$ and the second term is
appropriate for $H$ with central extension $(k-2h)$. Defining the new fields
$g'=h_-^{-1}gh_+,\ h'=h_-^{-1}h_+,\ h''=h_-$ and taking advantage of the
properties of the Haar measure, we can rewrite the measure and action in
decoupled form $Dg'\ Dh'\ Dh''$ and $S=-kI_0(g')+(k-2h)I_0(h')$. This
decoupled form emphasizes the close connection to the WZW path integral, and
gives us a clue for how to guess the effective quantum action.

However, $g',h'$ are not really decoupled, since we must consider sources
coupled to the {\it original fields}. Indeed, to derive the quantum effective
action one must introduce source terms and perform a Legendre transform. Since
these coupled $g',h',h''$ integrations are not easy to perform, we will
guess \BST\ the answer based on the remarks above and then try to justify it.
By analogy to \ewzw\ we suggest that the quantum effective action is given by
simply shifting $(-k)$ to $(-k+g)$ and $(k-2h)$ to $(k-2h)+h=k-h$.

\eqn\egwzw{ S^{eff}_{GWZW}=(-k+g)I_0(h_-^{-1}gh_+)-(-k+h)I_0(h_-^{-1}h_+) \ .}
This may now be rewritten back in terms of {\it classical} fields $g,A_+,A_-$
by using the definitions given before. We obtain

\eqn\effgwzw{ \eqalign {
     & S^{eff}_{GWZW}=(-k+g)\bl [I_0(g)+I_1(g,A_+,A_-)
                         +{g-h\over -k+g}I_2(A_+,A_-)\br ] \ , \cr
 &I_2(A_+,A_-)=I_3(A_+)+I_3(A_-)+{1\over 4\pi}\int d^2\sigma\ Tr(A_+A_-)\ . }}
 %
%
where we have defined $I_3(A_+)\equiv I_0(h_+)$,
$I_3(A_-)\equiv I_0(h_-^{-1})$ and used the
Polyakov-Wiegman formula \POLWIG\ to rewrite $I_0(h_-^{-1}h_+)\equiv
I_2(A_+,A_-)$ in the form above. Note that $I_2(A_+,A_-)$ is gauge invariant.
Our proposed effective action differs from the purely classical action \agwzw\
by the overall renormalization $(-k+g)$ and by the additional term
proportional to $(g-h)$. In the large $k$ limit (which is equivalent to small
$\hbar$) the effective quantum action reduces to the classical action, as it
should.

This is not yet the end of the story, because what we are really interested in
is the effective action for the sigma model after the gauge fields are
integrated out (and a unitary gauge fixed for $g$). In other words, sources
are not introduced for the original $A_\pm$, but only for $g$. The effect of
this is that the path integral over the above $A_\pm$ (or $h_\pm$) still needs
to be performed. At the outset, with the classical action, the path integral
over $A_\pm$ was purely gaussian, and therefore it could be performed
by simply
substituting the classical solutions for $A_\pm=A_\pm (g)$ back into the
action. This integration also introduces an anomaly which can be computed
exactly as a one loop effect. The anomaly gives the dilaton piece to be
added to the effective action

\eqn\adil{ S_{dil} \sim \int d^2\sigma \sqrt{\gamma}\ R^{(2)}(\gamma)\
\Phi(g)\ ,}
where $\gamma_{ab}, \gamma ,R^{(2)}$ are the metric, its determinant and
curvature on the worldsheet for any genus. In order to obtain the exact
dilaton we need to perform the $A_\pm$ integrals with the effective action,
not the classical one. However, in \effgwzw\ the parts $I_3(A_\pm)$ are
non-local in the $A_\pm$ (although they are local in $h_\pm$). The reason is
that $I_3(A_+)=I_0(h_+)\sim \int Tr(A_+\partial_-h_+h_+^{-1})+\cdots $, and we
cannot write $\partial_-h_+h_+^{-1}$ as a local function of $A_+$.
Furthermore, in the non-abelian case $I_3(A_\pm)$ have additional non-linear
terms. So, if we believe that the quantum effective action is indeed \effgwzw
, then the effective sigma model action we are seeking seems to be generally
non-local even in the abelian case (see also \TSEYT ). We will therefore
concentrate on just the zero modes. As shown below, we have managed to obtain
exactly the zero mode sector of the sigma model and proven that the geometry
does indeed reproduce correctly the exact geometry derived before
in the Hamiltonian formalism \BSexa\SFET. This is our justification
for \effgwzw .


\newsec{The zero mode sector}

To restrict ourselves to the zero mode sector we do dimensional reduction
by taking all the fields as functions of only $\tau$ (i.e. worldline rather
than worldsheet). This extracts the low energy point particle content of the
string. This technique proved to be very useful in the analysis of the GWZW
model at the classical limit \BSglo\ and we now use it for the conformally
exact action. The derivatives $\partial_\pm$ get replaced by $\partial_\tau$
and $A_\pm$ get replaced by $a_\pm=\partial_\tau h_\pm h^{-1}_\pm $. Then all
non-local and non-linear terms drop out and we obtain the effective action in
the zero mode sector

\eqn\actionzero{\eqalign{S_{eff}=&{-k+g\over 4\pi}\int d\tau \
Tr({1\over 2}\pat g^{-1}\pat g+a_-\pat g g^{-1}-a_+g^{-1}\pat g
+a_-ga_+g^{-1}-a_+a_-)\cr
&-{g-h\over 8\pi} \int d\tau\ Tr(a_+-a_-)^2 \ ,\cr }}
This action is gauge invariant for
$\tau$-dependent gauge transformations $\Lambda (\tau)$. Most notably the path
integral over $a_\pm$ is now Gaussian, and this permits the elimination of
$a_\pm$ through the classical equations of motion

\eqn\class{ (D_+gg^{-1})_H={g-h\over k-g}(a_+-a_-),\qquad
               (g^{-1}D_-g)_H={g-h\over k-g}(a_+-a_-) , \ }
where we have defined the covariant derivatives $D_\pm$ on the worldline
$D_\pm g=\partial_\tau g-[a_\pm ,g]$ and
the subscript $H$ indicates a projection to the Lie algebra of the
subgroup $H$. The system of equations \class\ is linear
and algebraic in $a_{\pm}$ and therefore it can be easily solved.
To do that and for further convenience it is useful to introduce a set
of matrices $\{t_A\}$ in the Lie algebra of $G$
which obey $Tr(t_A t_B)=\eta_{AB}$, where the Killing metric $\eta_{AB}$ is
diagonal and normalized to have $\pm 1$ eigenvalues. The subset of matrices
belonging to the Lie algebra of the subgroup $H$ will be denoted by $\{t_a\}$
with lower case subscripts or superscripts. Then we define the following
quantities

\eqn\difi{\eqalign{&L^H=(g^{-1}\pat g)_H\ , \qq \cr
&R^H=(-\pat g g^{-1})_H\ , \qq \cr
&M_{ab}=Tr(t_a g t_b g^{-1}-t_a t_b) \cr}
\eqalign{&L^A_{\mu}\pat X^{\mu}=Tr(g^{-1}\pat g t^A)\cr
&R^A_{\mu}\pat X^{\mu}=-Tr(\pat g g^{-1} t^A)\cr
& \ ,\cr}}

\no
where $X^{\m},\ \m=0,1,\cdots,d-1$ are the $d=dim(G/H)$
parameters in $g$ that are left over after going to a unitary gauge for $g$.
Then the solution of \class\ for $a_{\pm}$ is

\eqn\sola{\eqalign{&a_+=
\bigl(M^TM-\l (M+M^T)\bigr)^{-1}\bigl(M^T R^H-\l (L^H+R^H)\bigr)\cr
&a_-=
\bigl(MM^T-\l (M+M^T)\bigr)^{-1}\bigl(M L^H-\l (L^H+R^H)\bigr)
\ ,\cr}}

\no
where $\l={g-h\over k-g}$.
Substitution of these expressions back into \actionzero\ gives

\eqn\apoint{ S^{eff}_{point} = {k-g\over \pi}
\int d\tau\ G_{\m\n}\pat X^{\m} \pat X^{\n}\ ,  }

\no
where the metric $G_{\m\n}$ is defined as follows

\eqn\metric{\eqalign{G_{\m\n}=
&g_{\m\n}
+{1\over 8}\bigl([M^T M-\l (M+M^T)]^{-1}(M^T-\l I)\bigr)_{ab}
L^a_{\{\m}R^b_{\n\}} \cr
&-{1\over 8}\l \bigl(M^TM-\l (M+M^T)\bigr)^{-1}_{ab}L^a_{\m}L^b_{\n}\cr
&-{1\over 8}\l \bigl(MM^T-\l (M+M^T)\bigr)^{-1}_{ab}R^a_{\m}R^b_{\n}\ ,\cr}}

\no
with $g_{\m\n}$ being the part of the metric due to the kinetic
first term in $I_0(g)$

\eqn\smetric{g_{\m\n}=L^A_{\m}L^B_{\n}\eta_{AB}=R^A_{\m}R^B_{\n}\eta_{AB}\ ,}

\no
and where the curly brackets denote symmetrization with respect to
the appropriate indices.

We will illustrate applications of the above general result for several
abelian and non-abelian cosets. The simplest case is $SL(2,\IR)/\IR$,
but since this can be presented as a limit of more complicated cases,
we will give
the results for it after discussing others. This provides a check of
our methods.

Let us specialize to the three dimensional non-abelian coset
$SO(2,2)/SO(2,1)$ whose exact metric and dilaton was found in \BSexa\
with the Hamiltonian approach. We will find the metric in a patch of the
manifold corresponding, in the notation of \BSglo\BSexa\ to
$b=\cosh 2r\ ,u=\sin^2\th (\cosh 2t-1)\ ,v=\cosh 2t +1$, where $\{b,u,v\}$
are the global coordinates which cover the entire manifold.
The set of three matrices $\{t_a\}$ in the subgroup $H=SO(2,1)$ is

\eqn\mat{t_{01}={1\over \sqrt{2}} \pmatrix{0&1&0\cr 1&0&0\cr 0&0&0\cr}\ ,\
         t_{02}={1\over \sqrt{2}} \pmatrix{0&0&1\cr 0&0&0\cr 1&0&0\cr}\ ,\
         t_{12}={1\over \sqrt{2}} \pmatrix{0&0&0\cr 0&0&1\cr 0&-1&0\cr}\ .}

\no
The columns of the matrices $L_\mu^{\ a} ,R_\mu^{\ a}$ may be given as vectors
$L_{\m},R_{\m}\ ,\m=t,\th,r$

\eqn\LR{\eqalign{
&L_t=\sqrt{2}
\pmatrix{2c_r-2s_{\th}^2(c_r-1)\cr 2s_{\th}c_{\th} (c_r-1)\cr 0\cr}\ ,
\qq \cr
&L_{\th}=\sqrt{2}\pmatrix{0\cr 0\cr 1-c_r\cr}\ ,\qq \cr
&L_r=0\ ,\qq \cr}
\eqalign{&R_t=\sqrt{2}\pmatrix{-2\cr 0\cr 0\cr} \cr
&R_{\th}=\sqrt{2}\pmatrix{0\cr s_t(1-c_r)\cr c_t(1-c_r)\cr} \cr
&R_r=0 \cr}\ ,}

\no
and similarly the matrix $M_{ab}$ is

\eqn\Mab{(M_{ab})=\pmatrix{c_{\th}^2(c_r-1)&s_{\th} c_{\th}(c_r-1)&0\cr
c_ts_{\th}c_{\th}(c_r-1)&c_t(c_{\th}^2+s_{\th}^2c_r)-1&s_tc_r\cr
-s_ts_{\th}c_{\th}(c_r-1)&-s_t(c_{\th}^2+s_{\th}^2 c_r)&1-c_tc_r\cr}\ ,}

\no
where $c_r=\cosh 2r\ ,c_{\th}=\cos \th\ ,c_t=\cosh 2t$ and
$s_r=\sinh 2r\ ,s_{\th}=\sin \th , s_t=\sinh 2t$.
The non-zero components of the matrix $g_{\m\n}$ are
$g_{tt}=g_{rr}=1\ ,g_{\th\th}=(c_r-1)/2$.

\no
Using \metric-\Mab\ the non-zero components of the metric \metric\
take the following form

\eqn\matricp{\eqalign{&G_{rr}=1\cr
&G_{tt}=\b\bigl(\tanh^2 r\ \coth^2 t\ \tan^2\th -\coth^2 r\ {1\over \cos^2\th}
-{1\over k-1}\ {1\over \cos^2\th\ \sinh^2 t}\bigr)\cr
&G_{\th\th}=\b\bigl(\tanh^2 r -{1\over k-1}\ {1\over \cos^2\th}\bigr)\cr
&G_{t\th}=\b\ \tanh^2 r\ \coth t\ \tan\th\ ,\cr}}

\no
where the function $\b(r,t,\th)$ is defined as follows

\eqn\betab{\b^{-1}=1-{1\over k-1}\bigl({\coth^2 r \over \cos^2\th}
-\tanh^2 r({1\over \sinh^2 t}+\coth^2 t\ \tan^2\th)\bigr)
-{1\over (k-1)^2}\ {1\over \cos^2\th\ \sinh^2 t}\ .}

\no
It is not hard to check that the expression for the metric \matricp\ is
the same as the one we found in \BSexa\ with the Hamiltonian approach.
\foot{To
compare one should change variables from $(b,u,v)\to (t,\th,r)$
according to the prescription above.}


\newsec {The exact axion }

To obtain the axion $B_{\mu\nu}$ we need to retain the $\partial_\pm$ on
the worldsheet and then read off the coefficient
of ${1\over 2}(\pam X^\mu \pap X^\nu - \pam X^\nu \pap X^\mu )B_{\mu\nu}(X)$.
As already explained above we cannot do this fully
because of the non-local terms and non-abelian non-linearities, but we can
still obtain the axion as follows. We formally replace the $R^H,L^H$
in the expressions for $a_\pm$ and elsewhere by $R^H_\pm,L^H_\pm$,
where $R^H_\pm=(-
\partial_\pm gg^{-1})_H$ and $L^H_\pm=(g^{-1}\partial_\pm g)_H$.
We justify this step
by the conformal transformation properties for left and right movers. We then
substitute these forms of $A_\pm$ back into the action \effgwzw\ and extract
the desired axion from the quadratic part (which is local and a partner of the
metric). The expression we find for the axion $B_{\m\n}(X)$ is

\eqn\axion{B_{\m\n}=b_{\m\n}
+{1\over 8}\bigl([M^T M-\l (M+M^T)]^{-1}(M^T-\l I)\bigr)_{ab}
L^a_{[\m}R^b_{\n]}\ ,}

\no
where $b_{\m\n}$ is the part of the axion due to the Wess-Zumino term in
$I_0(g)$ and the brackets denote symmetrization with respect to the
appropriate indices.

In the particular case of the $SO(2,2)/SO(2,1)$ coset model we have found
\BSthree\ that for the semiclassical geometry ($k\to \infty$) the axion
field vanishes. However, when $k$ is finite we obtain a non-vanishing result,
which is given by the following expression

\eqn\axnonab{B_{t\th}={\b\over 2(k-1)} \tanh^2 r\ \coth t\ \tan\th\ ,}

\no
with the rest of the components being zero.
In terms of the global coordinates $\{b,u,v\}$ the corresponding expression
is

\eqn\axglo{B_{vu}={\b\over 8(k-1)}\ {b-1\over b+1}\ {1\over (v-2)(v-u-2)}\ .}

In section 7 we will obtain the exact axion for the three dimensional
black string model discussed in the semiclassical limit $k\to \infty$ in
(2nd ref. in \groups) and for any $k$ in \SFET.


\newsec{The exact dilaton}

To obtain the exact dilaton we must compute the anomaly in the integration
over $A_\pm$. However, as it was the case with the metric and the axion,
the local part of the dilaton can be obtained by going to the point
particle limit.
The effective action \actionzero\ contains a quadratic part in the gauge
fields which can be rewritten as follows

\eqn\actionn{{-k+g\over 4\pi}
\int d\tau\ Tr\bigl(a_-(M-\l I)a_+ +{\l \over 2}(a_- ^2 +a_+ ^2) \bigr)\ .}

\no
Integrating out the gauge fields $a_{\pm}$ gives a determinant that produces
the exact dilaton by identifying, ${\rm determinant}=e^{\Phi}$, that is

\eqn\dilaat{\Phi(X)=
\ln \bigl(\det(M)\sqrt{\det[I-\l (M^{-1}+(M^T)^{-1})]}\ \bigr)+{\rm const.}\ .}

\no
As an example, for the non-abelian coset $SO(2,2)/SO(2,1)$ this gives

\eqn\dilatt{\Phi=
\ln\bigl({\sinh^2 2r\ \sinh^2 t\ \cos^2 \th\over {\sqrt{\b}}}\bigr)
+{\rm const.}\ ,}

\no
or in terms of the global coordinates $\{b,u,v\}$

\eqn\ddil{\Phi=\ln\bigl({(b^2-1)(v-u-2)\over {\sqrt{\b}}}\bigr)
+{\rm const.}\ ,}

\no
which is exactly the expression found in \BSexa\ with the Hamiltonian approach.

We can use the general expressions for the exact metric \metric\
and dilaton \dilaat\ to check a theorem which we suggested before \BSexa .
We noticed
sometime ago \BSthree\ that the combination $e^\Phi \sqrt{-G}$ that appears in
the Laplacian \laplacian\ is actually independent of $k$. We had first
conjectured this by noting that, in the large $k$ limit, we could write this
quantity as the product of the Haar measure for $g$ times the Faddeev-Popov
determinant for fixing a unitary gauge for any $G/H$ gauged WZW model
 \foot {For a related statement for $SL(2,\IR)/\IR$ see also  \KIR ,
 and for clarifications see \TSEYT . In our previous work \BSthree\BShet\ we
erroneously stated that the path integral for the GWZW model requires an extra
gauge invariant factor $F(g)$ in the measure. Our error was due to the
omission of an anomaly factor. The correct measure at the outset is the Haar
measure for $g$ times the naive measure for the gauge fields $A_\pm$, and
$F=1$. This correction does not alter our theorem.
We thank E. Kiritsis and A.A. Tseytlin for comments on this point. }

\eqn\theorem{ e^\Phi \sqrt{-G} = \ {\rm (Haar) \times (Faddeev-Popov)}\ .   }
Both $G_{\mu\nu}$ and $\Phi$ receive $1/k$ corrections. But, by noting that
the right hand side is purely group theoretical we first conjectured that the
combination $e^\Phi\sqrt{-G}$ must remain $k$-independent. In our later
work for several non-abelian cases \BSexa\ we verified that this conjecture is
indeed true. Therefore, we stated the following theorem

\eqn\theoremm{ e^\Phi \sqrt{-G}\ ({\rm any}\ k)\ =\ e^\Phi \sqrt{-G}\ ({\rm
at}\ k=\infty )\ . }
We can reinforce this result by making additional observations.
First the path integral reasoning that allowed us to observe \theorem\
is equally valid when the effective action \effgwzw\
is used in place of the classical action \agwzw. Since the right hand side
of \theorem\
is purely group theoretical, \theorem\ should be valid both for the exact and
classical $G_{\m\n}$ and $\Phi$.
Since we have already computed the exact metric and dilaton one is now in
principle in a position to check the relation \theoremm\ in general.
However, the algebra required to compute $\sqrt{-G}$ is hard.
Instead, the
result for all cases relevant to strings in four dimensional curved spacetime
has already been computed explicitly in our previous papers and indeed for
abelian and non-abelian cases the theorem \theoremm\ is true.

To include the effects of
the dilaton we must add one more piece to the effective sigma model action

\eqn\fullaction{S^{eff}_{total}= S^{eff}_{sigma} + S^{eff}_{dil} \ ,}
where $S^{eff}_{dil}$ has the same form as \adil\ but with the exact dilaton
replacing the perturbative one. Here we have discussed mainly the zero mode
part of the total effective string action. The effective action for the
higher modes that follows from \egwzw\ and \effgwzw\ is generally non-local.

\newsec {Geodesics in the exact geometry}

In \BSglo\ the string (or particle) coordinates were defined as certain gauge
invariant combinations of the group parameters in $g$. In a specific unitary
gauge these invariants are related to the gauge fixed form of $g$ that defines
the string coordinates. Using this formalism a group theoretical method for
obtaining the solution to the geodesic equation was found and used to obtain
the geodesics in the classical geometry. It was shown that the solution to the
geodesic equation, which generally are complicated non-linear differential
equations for the string coordinates and hard to solve directly,
could be obtained by first solving the
equations of motion of the original variables $g(\tau),a_\pm(\tau)$ (which is
easy) and then forming the gauge invariant combinations {\it from the
solutions} for the group parameters in $g(\tau)$. We now apply the same method
to solve the geodesic equations in the exact geometry. So, we seek a solution
to the classical equations of motion given by \class\ and

\eqn\classs{ D_-(D_+gg^{-1})=\partial_\tau (a_--a_+)+[a_-,a_+] \ ,}
which follows from varying $g$, and
where $D_\pm$ have the same meaning as in \class . The method for solving
these equations is identical to \BSglo\ and the solution as a function of
proper time $\tau$ is

\eqn\solu{ g(\tau)=exp \bl ({k-g\over k-h}\ \alpha\ \tau\br ) \ g_0
\ exp \bl((P-\alpha)\ \tau \br)\ , \qq
\bl[ g_0 (P-\a) g_0^{-1}\br ]_H + \alpha =0 \ ,}
where $\alpha , P$ are constant matrices in the Lie algebra of $H$ and $G/H$
respectively, and $g_0$ is a constant group element. These matrices, which are
constrained by the second equation in \solu\ define the initial conditions for
any geodesic at $\tau=0$. The line element evaluated at this general solution
becomes

\eqn\lineel{ \bl ( {ds\over d\tau}\br )^2 = {k-g\over 8\pi} \
 Tr \bl (P^2+{g-h\over k-g}\ \alpha^2\br ) \ .}
The sign of this quantity determines whether the geodesic is timelike,
spacelike or lightlike, and it can be chosen {\it a
priori} as an initial
condition. The large $k$ analysis of \solu\ was given in \BSglo . With the new
$k$-dependence, and using the same methods as \BSglo ,
we have checked in a few
specific cases that the geodesic equations for the exact metric are indeed
solved with this group theoretical technique.


\newsec {Axial gauging and the \gh\ models}

So far we have concentrated on the vector gauging of WZW models.
For the axial gauging the subgroup $H$ should be abelian with zero Coxeter
number. The action is given by \agwzw\ but with $I_1(g,A_+,A_-)$ replaced by

\eqn\Iax{I^{\rm axial}_1(g,A_+,A_-)={1\over 4\pi}\int_M Tr(A_- \pap g g^{-1}
+A_+ g^{-1}\pam g -A_-gA_+g^{-1}-A_+A_-)\ .}

\no
Then if $A_{\pm}=-\papm h_{\pm}h_{\pm}^{-1}$
the analog of \effgwzw\ is

\eqn\effgwzwa{
S^{eff,axial}_{GWZW}=(-k+g)\bl [I_0(g)+I^{\rm axial}_1(g,A_+,A_-)
                         +{g\over -k+g}I_0(h_-^{-1}h_+)\br ] \ .}

\no
Let us specialize to the \gh\ coset models. For $d=3$ the semiclassical
aspects of the model were worked out in the 2nd ref. in \groups, for $d=4$ in
the 5th ref. in \groups\ and for general $d$ in \GIN. The conformally exact
geometry was found in \SFET\ with the Hamiltonian approach.
It is convenient to parametrize
the group element of $G=SL(2,\IR)\otimes SO(1,1)^{d-2}$ as follows

\eqn\grel{g=\pmatrix{g_0&0&\cdots&0\cr
                     0&g_1&\ldots&0\cr
                     \vdots&\vdots&\ddots&\vdots\cr
                     0&0&\cdots&g_{d-2}\cr}\ ,}

\no
where

\eqn\grslr{g_0=\pmatrix{a&u\cr-v&b\cr}, \qq ab+uv=1}

\no
and

\eqn\gru{g_i=\pmatrix{\cosh 2r_i&\sinh 2r_i\cr
\sinh 2r_i&\cosh 2r_i\cr}, \qq i=1,2,\cdots , d-2\ .}

\no
The infinitesimal generators for $SL(2,\IR)$ are

\eqn\gensl{j_0={q_0\over 2}\pmatrix{1&0\cr 0&-1\cr},\quad
j_+={q_0}\pmatrix{0&1\cr 0&0\cr},\quad
j_-={q_0}\pmatrix{0&0\cr -1&0\cr}}

\no
and those for the $SO(1,1)$'s

\eqn\genu{j_i={q_i}\pmatrix{0&1\cr 1&0\cr}, \qq i=1,2,\cdots , d-2\ .}

\no
The coefficients $q_i$ parametrize the embedding of $H=SO(1,1)$ into the
factored $SO(1,1)$'s in $G$ and are normalized to $\sum_{i=0}^{d-2}q_i ^2=1$.
The subgroup elements $h_{\pm}$ are parametrized in terms of two variables
$\phi_{\pm}$ as follows

\eqn\hpm{h_{\pm}=e^{-{1\over q_0}J_{U(1)} \phi_{\pm}}\ ,}

\no
where
\eqn\sgrel{J_{U(1)}=\pmatrix{j_0&0&\cdots&0\cr
                     0&j_1&\ldots&0\cr
                     \vdots&\vdots&\ddots&\vdots\cr
                     0&0&\cdots&j_{d-2}\cr}\ .}

\no
If we define two new variables $\phi=\phi_- -\phi_+$,
$\tilde \phi=\phi_- +\phi_+$ then in the gauge $b=\pm a$ the action
\effgwzwa\ takes the following form

\eqn\gaxial{\eqalign{S^{eff,axial}_{GWZW}={k'\over 16\pi}\int
&{1\over uv-1} \bigl(\pap(uv)\pam(uv)-2(uv-1)(\pap u\pam v+
 \pap v\pam u)\bigr)\cr
&+4{k\over k'}\sum_{i=1}^{d-2}\k_i \pap r_i \pam r_i \cr
&+(u\pap v-v \pap u-2 {k\over k'} \sum_{i=1}^{d-2}\k_i \eta_i \pap r_i)
 (\pam \phi +\pam \tilde \phi)\cr
&+(u\pam v-v \pam u+2 {k\over k'} \sum_{i=1}^{d-2}\k_i \eta_i \pam r_i)
 (\pap \phi -\pap \tilde \phi)\cr
&+(uv-1-{k\over k'} \r^2-{2\over k'})\pap\phi \pam\phi\cr
&+(1-uv+{k\over k'} \r^2)(\pap\tilde \phi\pam\tilde \phi
+\pam \phi \pap \tilde\phi-\pap \phi \pam \tilde\phi)\ .\cr}}

\no
where $k'=k-2$ is the renormalized value for the central extension
$k$ and $\eta_i\equiv q_i/q_0$, $\k_i\equiv k_i/k$,
$\r^2\equiv \sum_{i=1}^{d-2} \eta_i^2 \k_i$.

To extract the effective string model we now need to integrate out
$\phi$  and $\tilde\phi$, which is equivalent to integrating out $A_\pm$.
As discussed before this gives non-local contributions. Therefore, we may
again concentrate on the zero modes by dimensional reduction. Furthermore,
as discussed in section 4, we may restore formally $\partial_\tau \to
\partial_\pm$ in order to compute the axion. In some sense this procedure
extracts the local part of the effective action and preserves gauge invariance
with respect to $\tau$-dependent gauge transformations $\Lambda (\tau)$.
In fact, the local part of the effective action is an ambiguous notion and
the principle of $\tau$-dependent gauge invariance resolves this ambiguity
\foot{Our gauge invariant results differ in general from the local part
discussed in \TSEYT\ which is not gauge invariant with respect to $\Lambda
(\tau)$. Our form is required to produce the correct geometry that
agrees with the algebraic results. However, for the special case
$SL(2,\IR)/\IR$ the results for
the metric and dilaton agree accidentally with \TSEYT .}.
The upshot of these steps boils down to keeping
the local part of the solution of the classical equations
for the gauge fields $\phi$ and $\tilde \phi$

\eqn\gloc{\eqalign{&\papm \phi{\big|}_{\rm local}={v\papm u-u\papm v
\over uv-1-{k\over k'}\r^2-{2\over k'}}\cr
&\papm\tilde \phi\big|_{\rm local}=-2{k\over k'}
{\sum_{i=1}^{d-2}\k_i \eta_i \papm r_i
\over uv-1-{k\over k'}\r^2}\ .\cr}}

\no
Substitution of the above expressions into the action \gaxial\ gives
the following expression for the local part of the effective action

\eqn\faction{\eqalign{S^{local}_{eff}=&{k'\over 4\pi}\int
{1\over {k'/k\ (uv-1)-\r^2-2/k}}\bigl[-
{\r^2+2/k\over 4(uv-1)}\pap(uv)\pam(uv) \cr
 &+{1+\r^2\over 2}(\pap u\pam v+\pam u \pap v)\bigr]
 +{k\over k'}\sum_{i,j=1}^{d-2}\k_i(\d_{ij}
+{\eta_i\eta_j\k_j\over {k'/k\ (uv-1)-\r^2}}) \pap r_i \pam r_j\cr
 &+{1/2 \over k'/k\ (1-uv)+\r^2}
\sum_{i=1}^{d-2}\bigl((u\pap v-v\pap u)\ \k_i \eta_i \pam r_i -
(u\pam v-v\pam u)\ \k_i \eta_i \pap r_i\bigr)\ .\cr}}

\no
The first two lines in the above expression define a metric which is
precisely that found in \SFET\ with the Hamiltonian approach.
The third line defines an antisymmetric tensor (axion). As in ref. \SFET\ it
is useful to diagonalize the metric. Since the procedure is
exactly the same we are not going to repeat it here.
The answer is that, only a three dimensional part of the metric is non-trivial,
and the rest corresponds to flat directions. The three dimensional non-trivial
part of the metric, which describes a black string,
has the following form \SFET

\eqn\trdm{ds_{3d}^2=-(1-{r_+ \over r})\ dt^2
+(1-{r_- -r_q \over {r-r_q}})\ dx^2 +
{k'\over {8r^2}}(1-{r_+ \over r})^{-1}(1-{r_- \over r})^{-1}\ dr^2\ ,}

\no
where \SFET\ $r_+=\sqrt{2/k'}\ (\r^2+1)\ C$, $r_-=\sqrt{2/k'}\ (\r^2+2/k)\ C$
and $r_q=2/k\sqrt{2/k'}\ C$ (for $C$ see below the expression for the dilaton).
For the axion and its field strength we obtain a new result: $B_{tr}=B_{xr}
=0$, and

\eqn\axion{\eqalign{&B_{tx}=\sqrt{r_- -r_q\over r_+}\ {r-r_+\over r-r_q}\cr
&H_{rtx}=\partial_r B_{tx}
=\sqrt{r_- -r_q\over r_+}\ {r_+ -r_q\over (r-r_q)^2}\ .\cr}}

\no
To obtain the dilaton one has to integrate out
$\phi$ and $\tilde \phi$ in \gaxial. Then one gets
for the conformally exact dilaton the expression which was found in \SFET

\eqn\dila{C e^{\Phi}=(1-uv)\sqrt{[1+\r^2
+(\r^2+2/k){uv\over {1-uv}}][1+\r^2-2/k+\r^2{uv\over {1-uv}}]}\ ,}

\no
where $C$ is an arbitrary constant. In the variables which diagonalize
the metric, the dilaton takes the following form \SFET

\eqn\trddil{\Phi={1\over 2}\ln \bl(r(r-r_q)\br)+{1\over 2}\ln k'\ .}

\no
Therefore an additional piece $S^{eff}_{dil}(\Phi)$ must added to the action
in \faction.
The expressions for the metric, the dilaton, the axion and its field
strength tend to their semiclassical values (see 2nd ref. in \groups) in
the $k\to \infty$ limit, because then $r_q \to 0$.

It would be interesting to check that the expressions we found
for the metric, the axion and dilaton in this simple abelian model
indeed satisfy the perturbative
equations for conformal invariance beyond the 1-loop approximation.
For large $k$ the backgrounds of the $(2d\ {\rm black\ hole})\otimes \IR$
and the $3d$ black string are related by a duality transformation
as it was shown in \HOHOS.
Knowing the exact backgrounds (any $k$) for both geometries, may shed
some light into the form of the duality transformation beyond the leading
order in $\a'\sim1/k$.


\newsec{The $SL(2,\IR)/\IR$ model}

Since the simplest case $SL(2,\IR)/\IR$ is just a limit of the previous case
we will briefly derive in this section all the well known results.
In order to specialize the action \faction\ to
the case of the $SL(2,\IR)_{-k}/\IR$ model one should take $k_i=0$.
It follows that $\k_i=\r^2=0$ and the action \faction\ and dilaton
\dila\ take the following form

\eqn\actiontd{S^{eff}_{local}={k\over 8\pi}\int
{1\over {uv-1-2/k'}}\bigl(-
{1/k\over uv-1}\pap(uv)\pam(uv) +(\pap u\pam v+\pam u \pap v)\bigr)
+S^{dil}_{2d}(\Phi)\ ,}

\no
and

\eqn\dilat{C' e^{\Phi}=(1-uv)\sqrt{1-{2\over k}{uv\over uv-1}}\ ,}

\no
where $C'$ is a constant related to $C$ in \dila.
In the region where $uv>1$ we change variables from $(u,v)\to (t,r)$
as follows

\eqn\chan{u=\cosh r\ e^t\ ,\quad v=\cosh r\ e^{-t}\ .}

\no
Then the action \actiontd\ and the dilaton \dilat\ can be written as

\eqn\act{\eqalign{&S^{local}_{2d}={k'\over 4\pi}\int \pap r\pam r-
f(r)\pap t\pam t +S^{dil}_{2d}(\Phi)\cr
&\Phi=\ln(\sinh 2r/f(r))+{\rm const.}\ ,\cr}}

\no
where $f(r)=1/(\tanh^2 r-2/k)$, thus reproducing the exact
expressions for the metric and dilaton of the $2d$ black hole as they
were computed in \DVV\BSexa. One could also use the effective action
appropriate for vectorial gauging \effgwzw\ to obtain all of the results
in this section.


\newsec{ Conclusion}

We have suggested the form of the effective quantum action for the general
Abelian or non-Abelian GWZW model, and verified that it works, at least in the
zero mode sector. Furthermore, we have obtained new general results for the
conformally exact axion field and geodesics. The zero mode sector determines
the point particle behavior of the underlying string theory and is the only
part relevant for the low energy physics. Therefore, although our methods have
yielded incomplete results for the full string theory, they are adequate to
extract the most relevant physical information on the curved spacetime
geometry. Based on the agreement with the algebraic Hamiltonian approach in
the zero mode sector, we conjecture that, before the integration over $A_\pm$,
the forms \egwzw\effgwzw\ may be trusted for all the higher modes as well.

\newsec {Acknowledgements}

We thank A.A. Tseytlin for very useful discussions and collaboration.
We also thank the Aspen Center for Physics where part of this work was done.

\listrefs
\end